\documentclass[11pt,a4paper]{article}

\usepackage{jheppub}
\usepackage{xcolor}
\usepackage[normalem]{ulem}
\usepackage[pass]{geometry}
\usepackage{float}
\usepackage{longtable}
\usepackage{array}
\usepackage{booktabs}
\usepackage{arydshln}
\usepackage{bm}
\usepackage{placeins}
\hypersetup{
  colorlinks=true,
  linkcolor=black,
  citecolor=black,
  urlcolor=black,
  hypertexnames=false,
  pdftitle={A singlet framework for the mu problem and type-II seesaw in a generalized NMSSM},
  pdfauthor={Xiaokang Du, Yusi Pan and Fei Wang}
}

\newcommand{\msusy}{m_{\rm SUSY}}
\newcommand{\mew}{m_{\rm EW}}
\newcommand{\mueff}{\mu_{\rm eff}}
\newcommand{\Tr}{\operatorname{Tr}}
\newcommand{\BMvalue}[2]{\ifcsname BM@#1@#2\endcsname\csname BM@#1@#2\endcsname\else\textemdash\fi}
\def\BMbinocase{B0}
\newcommand{\BMrow}[2]{#1 & \BMvalue{\BMcaseA}{#2} & \BMvalue{\BMcaseB}{#2}\\}
\newcommand{\BMgroup}[1]{\multicolumn{3}{@{}l}{\textit{#1}}\\}
\newcommand{\BMsize}{\linespread{1}\fontsize{9}{10.2}\selectfont}
\newcommand{\BMtable}[4]{%
\def\BMcaseA{#1}\def\BMcaseB{#2}%
\begingroup\BMsize\setlength{\LTcapwidth}{\textwidth}\setlength{\tabcolsep}{4pt}\renewcommand{\arraystretch}{1.07}%
\begin{longtable}{@{}>{\raggedright\arraybackslash}p{.34\textwidth}*{2}{>{\raggedright\arraybackslash}p{.30\textwidth}}@{}}
\toprule \BMquantity & BP-#1 & BP-#2\\\midrule\endfirsthead
\toprule \BMquantity & BP-#1 & BP-#2\\\midrule\endhead
\bottomrule\endfoot
\bottomrule
\caption{#3}\label{#4}\\
\endlastfoot
\BMgroup{\BMinput}
\BMrow{$m_0(M_T)$ [GeV]}{m0}
\BMrow{$M_{1,2,3}(M_T)$ [GeV]}{gauginos}
\BMrow{$M_S(M_T)$ [GeV]}{MS}
\BMrow{$\tan\beta$}{tanbeta}
\BMrow{$\lambda(Q_{\rm SUSY}),\ \lambda_T^{\rm in}$}{lambda_lambdaT}
\BMrow{$v_S(Q_{\rm SUSY}),\ M_T$ [GeV]}{vS_MT}
\BMrow{$\mu_0(M_T)$ [GeV]}{mu0}
\BMrow{$A_0(M_T),\ A_\lambda(M_T)$ [GeV]}{A0_Alambda}
\BMrow{$B_{\mu_0}(M_T),\ B_{M_S}(M_T)$ [GeV$^2$]}{Bmu_BMS}
\BMrow{$\chi_u^{\rm in},\ \chi_l^{\rm in}$}{chi_u_l}
\BMgroup{\BMrunning}
\BMrow{$Q_{\rm SUSY}$ [GeV]}{Q}
\BMrow{$\mu_{\rm eff}(Q_{\rm SUSY})$ [GeV]}{mueff}
\BMrow{$\Delta_\mu(Q_{\rm SUSY})$}{Delta_mu}
\BMgroup{\BMhiggs}
\BMrow{$m_{h_{1,2,3}}$ [GeV]}{higgses}
\BMrow{$m_{a_{1,2}},\ m_{H^\pm_1}$ [GeV]}{pseudoscalar_charged}
\BMgroup{\BMspectrum}
\BMrow{$m_{\widetilde\chi^0_{1,2,3}}$ [GeV]}{neutralinos123}
\BMrow{$m_{\widetilde\chi^0_{4,5}}$ [GeV]}{neutralinos45}
\BMrow{$m_{\widetilde\chi^\pm_{1,2}}$ [GeV]}{charginos}
\BMrow{$f_B,\ f_W$}{fB_fW}
\BMrow{$f_H,\ f_S$}{fH_fS}
\BMrow{$m_{\widetilde g}$ [GeV]}{gluino}
\BMrow{$m_{\widetilde t_{1,2}}$ [GeV]}{stops}
\BMrow{$m_{\widetilde b_{1,2}}$ [GeV]}{sbottoms}
\BMrow{$m_{\widetilde\tau_{1,2}}$ [GeV]}{staus}
\BMgroup{\BMdarkmatter}
\BMrow{$\Omega_\chi h^2$}{omega}
\BMrow{$\sigma_{\rm SI}^{p,\mathrm{eff}},\ \sigma_{\rm SI}^{n,\mathrm{eff}}$ [pb]}{si_p_n}
\BMrow{$\sigma_{\rm SD}^{p,\mathrm{eff}},\ \sigma_{\rm SD}^{n,\mathrm{eff}}$ [pb]}{sd_p_n}
\BMrow{$R_{\rm SI}$}{R_SI}
\BMrow{$\delta_{\rm EWino},\ \delta_{\widetilde g}$}{gap_ewino_gluino}
\BMrow{\BMmechanism}{mechanism}
\BMrow{$w_{\chi\chi}$}{w_ann}
\BMrow{$w_{\rm coann}$}{w_coann}
\BMrow{\BMpolestates}{pole_states}
\BMrow{$\Gamma_\phi$ [GeV]}{pole_widths}
\BMrow{$(m_\phi-2m_\chi)/\Gamma_\phi$}{pole_offset}
\BMgroup{\BMconstraints}
\BMrow{$R_{\rm HB},\ p_{\rm HS}$}{HB_HS}
\BMrow{${\rm BR}(B\to X_s\gamma)$}{BR_bsgamma}
\BMrow{${\rm BR}(B_s\to\mu^+\mu^-)$}{BR_bsmumu}
\BMrow{${\rm BR}(\mu\to e\gamma)$}{BR_muegamma}
\BMrow{\BMLHC}{LHC_status}
\BMrow{$\|\chi_l\|_2$}{chi_l_norm}
\BMgroup{\BMreactions}
\BMrow{\BMreaction~1}{reaction1}
\BMrow{\BMreaction~2}{reaction2}
\ifx\BMcaseA\BMbinocase\else
 \BMrow{\BMreaction~3}{reaction3}
 \BMrow{\BMreaction~4}{reaction4}
 \BMrow{\BMreaction~5}{reaction5}
\fi
\end{longtable}\endgroup}

\expandafter\def\csname BM@B0@m0\endcsname{$1.07\times10^{4}$}
\expandafter\def\csname BM@B0@gauginos\endcsname{$1.09\times10^{3},\ -4.21\times10^{3},\ 6.85\times10^{3}$}
\expandafter\def\csname BM@B0@MS\endcsname{$1.57\times10^{3}$}
\expandafter\def\csname BM@B0@tanbeta\endcsname{$3.16$}
\expandafter\def\csname BM@B0@lambda_lambdaT\endcsname{$0.0173,\ 0.0655$}
\expandafter\def\csname BM@B0@vS_MT\endcsname{$3.17\times10^{14},\ 1.47\times10^{13}$}
\expandafter\def\csname BM@B0@mu0\endcsname{$-4.24\times10^{12}$}
\expandafter\def\csname BM@B0@A0_Alambda\endcsname{$1.41\times10^{3},\ 5.16\times10^{-8}$}
\expandafter\def\csname BM@B0@Bmu_BMS\endcsname{$5.90\times10^{11},\ -5.73\times10^{6}$}
\expandafter\def\csname BM@B0@chi_u_l\endcsname{$0.0503,\ 0.251$}
\expandafter\def\csname BM@B0@Q\endcsname{$12341.78$}
\expandafter\def\csname BM@B0@mueff\endcsname{$2217.42$}
\expandafter\def\csname BM@B0@Delta_mu\endcsname{$1.75\times10^{9}$}
\expandafter\def\csname BM@B0@neutralinos123\endcsname{$634.50,\ 1568.99,\ 2262.98$}
\expandafter\def\csname BM@B0@neutralinos45\endcsname{$2265.33,\ 3915.24$}
\expandafter\def\csname BM@B0@charginos\endcsname{$2264.03,\ 3915.21$}
\expandafter\def\csname BM@B0@fB_fW\endcsname{$0.99939231,\ 9.3040157\times10^{-8}$}
\expandafter\def\csname BM@B0@fH_fS\endcsname{$6.0759027\times10^{-4},\ 8.9136442\times10^{-9}$}
\expandafter\def\csname BM@B0@higgses\endcsname{$124.32,\ 1270.64,\ 1564.51$}
\expandafter\def\csname BM@B0@pseudoscalar_charged\endcsname{$1269.45,\ 3729.24,\ 1271.75$}
\expandafter\def\csname BM@B0@gluino\endcsname{$14338.51$}
\expandafter\def\csname BM@B0@stops\endcsname{$11159.36,\ 13647.39$}
\expandafter\def\csname BM@B0@sbottoms\endcsname{$13645.89,\ 15462.90$}
\expandafter\def\csname BM@B0@staus\endcsname{$10396.89,\ 11172.11$}
\expandafter\def\csname BM@B0@omega\endcsname{$0.101342$}
\expandafter\def\csname BM@B0@si_p_n\endcsname{$6.59\times10^{-12},\ 6.67\times10^{-12}$}
\expandafter\def\csname BM@B0@sd_p_n\endcsname{$3.59\times10^{-9},\ 2.83\times10^{-9}$}
\expandafter\def\csname BM@B0@R_SI\endcsname{$0.342$}
\expandafter\def\csname BM@B0@gap_ewino_gluino\endcsname{$1.47,\ 21.6$}
\expandafter\def\csname BM@B0@HB_HS\endcsname{$0.511,\ 0.571$}
\expandafter\def\csname BM@B0@BR_bsgamma\endcsname{$3.19\times10^{-4}$}
\expandafter\def\csname BM@B0@BR_bsmumu\endcsname{$3.16\times10^{-9}$}
\expandafter\def\csname BM@B0@chi_l_norm\endcsname{$0.271$}
\expandafter\def\csname BM@B0@BR_muegamma\endcsname{$\leq 6.57\times10^{-50}$}
\expandafter\def\csname BM@B0@w_ann\endcsname{$0.994239$}
\expandafter\def\csname BM@B0@w_coann\endcsname{$0.000000$}
\expandafter\def\csname BM@B0@mechanism\endcsname{\BMfunnel}
\expandafter\def\csname BM@B0@pole_states\endcsname{$h_2,\ a_1$}
\expandafter\def\csname BM@B0@pole_offset\endcsname{$0.287,\ 0.0777$}
\expandafter\def\csname BM@B0@pole_widths\endcsname{$5.74,\ 5.88$}
\expandafter\def\csname BM@B0@reaction1\endcsname{$\widetilde\chi^0_1\widetilde\chi^0_1\to t\bar t$\,$(0.974317)$}
\expandafter\def\csname BM@B0@reaction2\endcsname{$\widetilde\chi^0_1\widetilde\chi^0_1\to b\bar b$\,$(0.019922)$}
\expandafter\def\csname BM@B1@m0\endcsname{$1.02\times10^{4}$}
\expandafter\def\csname BM@B1@gauginos\endcsname{$4.04\times10^{3},\ 4.61\times10^{3},\ 1.54\times10^{3}$}
\expandafter\def\csname BM@B1@MS\endcsname{$4.67\times10^{3}$}
\expandafter\def\csname BM@B1@tanbeta\endcsname{$12.1$}
\expandafter\def\csname BM@B1@lambda_lambdaT\endcsname{$0.0806,\ 0.607$}
\expandafter\def\csname BM@B1@vS_MT\endcsname{$4.14\times10^{9},\ 1.78\times10^{9}$}
\expandafter\def\csname BM@B1@mu0\endcsname{$-2.48\times10^{8}$}
\expandafter\def\csname BM@B1@A0_Alambda\endcsname{$5.11\times10^{3},\ 0.0286$}
\expandafter\def\csname BM@B1@Bmu_BMS\endcsname{$1.25\times10^{9},\ -4.01\times10^{7}$}
\expandafter\def\csname BM@B1@chi_u_l\endcsname{$2.04\times10^{-6},\ 0.332$}
\expandafter\def\csname BM@B1@Q\endcsname{$8928.58$}
\expandafter\def\csname BM@B1@mueff\endcsname{$4341.55$}
\expandafter\def\csname BM@B1@Delta_mu\endcsname{$5.43\times10^{4}$}
\expandafter\def\csname BM@B1@neutralinos123\endcsname{$3064.69,\ 4358.84,\ 4384.45$}
\expandafter\def\csname BM@B1@neutralinos45\endcsname{$4510.21,\ 4659.22$}
\expandafter\def\csname BM@B1@charginos\endcsname{$4358.62,\ 4510.14$}
\expandafter\def\csname BM@B1@fB_fW\endcsname{$0.99933271,\ 8.5424045\times10^{-7}$}
\expandafter\def\csname BM@B1@fH_fS\endcsname{$6.6321815\times10^{-4},\ 3.2205769\times10^{-6}$}
\expandafter\def\csname BM@B1@higgses\endcsname{$125.42,\ 4645.39,\ 9402.40$}
\expandafter\def\csname BM@B1@pseudoscalar_charged\endcsname{$9402.36,\ 10076.36,\ 9402.83$}
\expandafter\def\csname BM@B1@gluino\endcsname{$3094.21$}
\expandafter\def\csname BM@B1@stops\endcsname{$8255.77,\ 9656.15$}
\expandafter\def\csname BM@B1@sbottoms\endcsname{$9655.25,\ 10438.65$}
\expandafter\def\csname BM@B1@staus\endcsname{$10279.19,\ 10407.32$}
\expandafter\def\csname BM@B1@omega\endcsname{$0.096738$}
\expandafter\def\csname BM@B1@si_p_n\endcsname{$5.46\times10^{-12},\ 5.52\times10^{-12}$}
\expandafter\def\csname BM@B1@sd_p_n\endcsname{$1.02\times10^{-9},\ 8.84\times10^{-10}$}
\expandafter\def\csname BM@B1@R_SI\endcsname{$0.0619$}
\expandafter\def\csname BM@B1@gap_ewino_gluino\endcsname{$0.422,\ 0.00963$}
\expandafter\def\csname BM@B1@HB_HS\endcsname{$0.602,\ 0.537$}
\expandafter\def\csname BM@B1@BR_bsgamma\endcsname{$3.16\times10^{-4}$}
\expandafter\def\csname BM@B1@BR_bsmumu\endcsname{$3.16\times10^{-9}$}
\expandafter\def\csname BM@B1@chi_l_norm\endcsname{$0.738$}
\expandafter\def\csname BM@B1@BR_muegamma\endcsname{$\leq 1.70\times10^{-32}$}
\expandafter\def\csname BM@B1@w_ann\endcsname{$0.000000$}
\expandafter\def\csname BM@B1@w_coann\endcsname{$0.999970$}
\expandafter\def\csname BM@B1@mechanism\endcsname{\BMgluino}
\expandafter\def\csname BM@B1@pole_states\endcsname{\textemdash}
\expandafter\def\csname BM@B1@pole_offset\endcsname{\textemdash}
\expandafter\def\csname BM@B1@pole_widths\endcsname{\textemdash}
\expandafter\def\csname BM@B1@reaction1\endcsname{$\widetilde g\widetilde g\to q\bar q$\,$(0.631590)$}
\expandafter\def\csname BM@B1@reaction2\endcsname{$\widetilde g\widetilde g\to gg$\,$(0.368380)$}
\expandafter\def\csname BM@S0@m0\endcsname{$1.57\times10^{3}$}
\expandafter\def\csname BM@S0@gauginos\endcsname{$1.64\times10^{3},\ 5.60\times10^{2},\ 6.58\times10^{3}$}
\expandafter\def\csname BM@S0@MS\endcsname{$386$}
\expandafter\def\csname BM@S0@tanbeta\endcsname{$6.07$}
\expandafter\def\csname BM@S0@lambda_lambdaT\endcsname{$0.0565,\ 0.0986$}
\expandafter\def\csname BM@S0@vS_MT\endcsname{$5.19\times10^{10},\ 3.62\times10^{9}$}
\expandafter\def\csname BM@S0@mu0\endcsname{$-2.18\times10^{9}$}
\expandafter\def\csname BM@S0@A0_Alambda\endcsname{$1.34\times10^{3},\ -1.40\times10^{-4}$}
\expandafter\def\csname BM@S0@Bmu_BMS\endcsname{$4.76\times10^{8},\ 1.55\times10^{4}$}
\expandafter\def\csname BM@S0@chi_u_l\endcsname{$1.21\times10^{-6},\ 2.76$}
\expandafter\def\csname BM@S0@Q\endcsname{$7476.66$}
\expandafter\def\csname BM@S0@mueff\endcsname{$484.24$}
\expandafter\def\csname BM@S0@Delta_mu\endcsname{$4.28\times10^{6}$}
\expandafter\def\csname BM@S0@neutralinos123\endcsname{$385.15,\ 457.23,\ 499.73$}
\expandafter\def\csname BM@S0@neutralinos45\endcsname{$598.39,\ 1194.76$}
\expandafter\def\csname BM@S0@charginos\endcsname{$459.86,\ 599.89$}
\expandafter\def\csname BM@S0@fB_fW\endcsname{$1.0337625\times10^{-5},\ 6.6377422\times10^{-4}$}
\expandafter\def\csname BM@S0@fH_fS\endcsname{$0.0042406824,\ 0.99508521$}
\expandafter\def\csname BM@S0@higgses\endcsname{$126.24,\ 374.43,\ 776.49$}
\expandafter\def\csname BM@S0@pseudoscalar_charged\endcsname{$330.48,\ 775.67,\ 779.99$}
\expandafter\def\csname BM@S0@gluino\endcsname{$10523.62$}
\expandafter\def\csname BM@S0@stops\endcsname{$7226.43,\ 7735.52$}
\expandafter\def\csname BM@S0@sbottoms\endcsname{$7729.52,\ 8183.14$}
\expandafter\def\csname BM@S0@staus\endcsname{$1520.84,\ 1669.80$}
\expandafter\def\csname BM@S0@omega\endcsname{$0.127536$}
\expandafter\def\csname BM@S0@si_p_n\endcsname{$2.21\times10^{-12},\ 2.23\times10^{-12}$}
\expandafter\def\csname BM@S0@sd_p_n\endcsname{$3.46\times10^{-8},\ 2.65\times10^{-8}$}
\expandafter\def\csname BM@S0@R_SI\endcsname{$0.189$}
\expandafter\def\csname BM@S0@gap_ewino_gluino\endcsname{$0.187,\ 26.3$}
\expandafter\def\csname BM@S0@HB_HS\endcsname{$0.638,\ 0.681$}
\expandafter\def\csname BM@S0@BR_bsgamma\endcsname{$3.28\times10^{-4}$}
\expandafter\def\csname BM@S0@BR_bsmumu\endcsname{$3.15\times10^{-9}$}
\expandafter\def\csname BM@S0@chi_l_norm\endcsname{$2.58$}
\expandafter\def\csname BM@S0@BR_muegamma\endcsname{$\leq 1.49\times10^{-31}$}
\expandafter\def\csname BM@S0@w_ann\endcsname{$0.992883$}
\expandafter\def\csname BM@S0@w_coann\endcsname{$0.000000$}
\expandafter\def\csname BM@S0@mechanism\endcsname{\BMSfunnel}
\expandafter\def\csname BM@S0@pole_states\endcsname{$h_3,\ a_2$}
\expandafter\def\csname BM@S0@pole_offset\endcsname{$4.58,\ 3.78$}
\expandafter\def\csname BM@S0@pole_widths\endcsname{$1.35,\ 1.42$}
\expandafter\def\csname BM@S0@reaction1\endcsname{$\widetilde\chi^0_1\widetilde\chi^0_1\to t\bar t$\,$(0.715136)$}
\expandafter\def\csname BM@S0@reaction2\endcsname{$\widetilde\chi^0_1\widetilde\chi^0_1\to b\bar b$\,$(0.225717)$}
\expandafter\def\csname BM@S0@reaction3\endcsname{$\widetilde\chi^0_1\widetilde\chi^0_1\to\tau^-\tau^+$\,$(0.040078)$}
\expandafter\def\csname BM@S0@reaction4\endcsname{$\widetilde\chi^0_1\widetilde\chi^0_1\to h_2 Z$\,$(0.011952)$}
\expandafter\def\csname BM@S0@reaction5\endcsname{\textemdash}
\expandafter\def\csname BM@S1@m0\endcsname{$1.47\times10^{3}$}
\expandafter\def\csname BM@S1@gauginos\endcsname{$2.32\times10^{3},\ 1.47\times10^{3},\ 6.21\times10^{3}$}
\expandafter\def\csname BM@S1@MS\endcsname{$1.44\times10^{3}$}
\expandafter\def\csname BM@S1@tanbeta\endcsname{$14.4$}
\expandafter\def\csname BM@S1@lambda_lambdaT\endcsname{$0.0189,\ 0.0359$}
\expandafter\def\csname BM@S1@vS_MT\endcsname{$1.22\times10^{11},\ 3.09\times10^{9}$}
\expandafter\def\csname BM@S1@mu0\endcsname{$-1.72\times10^{9}$}
\expandafter\def\csname BM@S1@A0_Alambda\endcsname{$1.20\times10^{3},\ 9.38\times10^{-4}$}
\expandafter\def\csname BM@S1@Bmu_BMS\endcsname{$1.54\times10^{8},\ -4.33\times10^{6}$}
\expandafter\def\csname BM@S1@chi_u_l\endcsname{$7.35\times10^{-6},\ 0.348$}
\expandafter\def\csname BM@S1@Q\endcsname{$7332.10$}
\expandafter\def\csname BM@S1@mueff\endcsname{$4567.87$}
\expandafter\def\csname BM@S1@Delta_mu\endcsname{$3.58\times10^{5}$}
\expandafter\def\csname BM@S1@neutralinos123\endcsname{$1436.52,\ 1441.72,\ 1699.66$}
\expandafter\def\csname BM@S1@neutralinos45\endcsname{$4595.13,\ 4595.97$}
\expandafter\def\csname BM@S1@charginos\endcsname{$1440.66,\ 4596.52$}
\expandafter\def\csname BM@S1@fB_fW\endcsname{$1.6415369\times10^{-6},\ 2.1590126\times10^{-4}$}
\expandafter\def\csname BM@S1@fH_fS\endcsname{$9.5422135\times10^{-7},\ 0.99978150$}
\expandafter\def\csname BM@S1@higgses\endcsname{$127.38,\ 1437.93,\ 4584.12$}
\expandafter\def\csname BM@S1@pseudoscalar_charged\endcsname{$3275.50,\ 4584.10,\ 4584.94$}
\expandafter\def\csname BM@S1@gluino\endcsname{$9933.65$}
\expandafter\def\csname BM@S1@stops\endcsname{$7174.52,\ 7492.15$}
\expandafter\def\csname BM@S1@sbottoms\endcsname{$7481.71,\ 7696.86$}
\expandafter\def\csname BM@S1@staus\endcsname{$1620.95,\ 1700.30$}
\expandafter\def\csname BM@S1@omega\endcsname{$0.121312$}
\expandafter\def\csname BM@S1@si_p_n\endcsname{$7.70\times10^{-17},\ 7.79\times10^{-17}$}
\expandafter\def\csname BM@S1@sd_p_n\endcsname{$1.00\times10^{-14},\ 7.30\times10^{-15}$}
\expandafter\def\csname BM@S1@R_SI\endcsname{$1.82\times10^{-6}$}
\expandafter\def\csname BM@S1@gap_ewino_gluino\endcsname{$0.00288,\ 5.92$}
\expandafter\def\csname BM@S1@HB_HS\endcsname{$0.731,\ 0.550$}
\expandafter\def\csname BM@S1@BR_bsgamma\endcsname{$3.16\times10^{-4}$}
\expandafter\def\csname BM@S1@BR_bsmumu\endcsname{$3.16\times10^{-9}$}
\expandafter\def\csname BM@S1@chi_l_norm\endcsname{$0.357$}
\expandafter\def\csname BM@S1@BR_muegamma\endcsname{$\leq 1.01\times10^{-34}$}
\expandafter\def\csname BM@S1@w_ann\endcsname{$0.000000$}
\expandafter\def\csname BM@S1@w_coann\endcsname{$0.963331$}
\expandafter\def\csname BM@S1@mechanism\endcsname{\BMEWino}
\expandafter\def\csname BM@S1@pole_states\endcsname{\textemdash}
\expandafter\def\csname BM@S1@pole_offset\endcsname{\textemdash}
\expandafter\def\csname BM@S1@pole_widths\endcsname{\textemdash}
\expandafter\def\csname BM@S1@reaction1\endcsname{$\widetilde\chi^0_2\widetilde\chi^-_1\to d_i\bar u_i$\,$(0.334441)$}
\expandafter\def\csname BM@S1@reaction2\endcsname{$\widetilde\chi^-_1\widetilde\chi^+_1\to q\bar q$\,$(0.183184)$}
\expandafter\def\csname BM@S1@reaction3\endcsname{$\widetilde\chi^-_1\widetilde\chi^-_1\to W^-W^-$\,$(0.100488)$}
\expandafter\def\csname BM@S1@reaction4\endcsname{$\widetilde\chi^0_2\widetilde\chi^0_2\to W^-W^+$\,$(0.090577)$}
\expandafter\def\csname BM@S1@reaction5\endcsname{$\widetilde\chi^0_2\widetilde\chi^-_1\to ZW^-$\,$(0.083666)$}

\def\BMrunning{Running quantities}
\def\BMhiggs{Higgs masses}
\def\BMconstraints{Auxiliary observables and constraints}
\def\BMLHC{LHC signal bounds (LO)}
\expandafter\def\csname BM@B0@LHC_status\endcsname{Bound below limits}
\expandafter\def\csname BM@B1@LHC_status\endcsname{Rate-based bound below limits}
\expandafter\def\csname BM@S0@LHC_status\endcsname{Bound below limits}
\expandafter\def\csname BM@S1@LHC_status\endcsname{Bound below limits}
\def\BMmechanism{Dominant thermal mechanism}
\def\BMpolestates{Relevant Higgs pole(s)}
\def\BMfunnel{$h_2/a_1$ Higgs funnel}
\def\BMSfunnel{$h_3/a_2$ Higgs funnel}
\def\BMgluino{Gluino coannihilation}
\def\BMEWino{Wino coannihilation}
\def\BMquantity{Quantity}
\def\BMinput{Input parameters}
\def\BMspectrum{Superpartner spectrum and LSP composition}
\def\BMdarkmatter{Dark matter and thermal mechanism}
\def\BMreactions{Leading freeze-out reactions (weight)}
\def\BMreaction{Reaction}

\title{A singlet framework for the $\mu$ problem and type-II seesaw in a generalized NMSSM}

\author[a,b]{Xiaokang Du}
\author[c]{Yusi Pan}
\author[d,1]{Fei Wang\note[1]{Corresponding author.}}

\affiliation[a]{Institute of Physics, Henan Academy of Sciences, Zhengzhou 450046, P. R. China}
\affiliation[b]{Centre for Theoretical Physics, Henan Normal University, Xinxiang 453007, P. R. China}
\affiliation[c]{School of Physics and Electronic Engineering, Shangqiu Normal University, Shangqiu 476000, P. R. China}
\affiliation[d]{School of Physics, Zhengzhou University, Zhengzhou 450000, P. R. China}

\emailAdd{xkdu@hnas.ac.cn}
\emailAdd{panyusi0406@foxmail.com}
\emailAdd{feiwang@zzu.edu.cn}

\abstract{
We study a generalized NMSSM in which a gauge singlet links the Higgs bilinear to the type-II seesaw scale. The singlet vacuum expectation value contributes to $\mu_{\rm eff}$ and generates the supersymmetric mass of a vectorlike triplet pair. Integrating out the triplets gives the Weinberg operator, and their gauge charges admit a possible messenger interpretation. Large cancellation between the bare $\mu_0$ term and the singlet VEV generated bilinear Higgs term that keeps $\mu_{\rm eff}$ near the soft scale is still natural in the supersymmetric limit (by the supersymmetric non-renormalization theorem), although the concrete value of $\mu_{\rm eff}$ also depends slightly on the soft SUSY breaking parameters and finite threshold matching. The studied bino-like and singlino-like neutralino solutions accommodate triplet input scales approaching $10^{15}$ GeV together with electroweak-to-TeV dark matter. Representative spectra consistent with heavy-Higgs funnels or coannihilation with gluinos or winos yield allowed relic abundances while satisfying direct-detection constraints.
}

\begin{document}
\maketitle

\clearpage

\section{Introduction}

The minimal supersymmetric standard model (MSSM) contains a supersymmetric Higgs bilinear $\mu H_uH_d$ allowed by the gauge symmetries. A viable electroweak vacuum without large cancellations requires $|\mu|$ to be comparable to the soft supersymmetry-breaking scale. Explaining this relation between a supersymmetric mass and the soft scale is the $\mu$ problem \cite{Kim:1983dt,Giudice:1988yz}. It motivates mechanisms that connect $\mu$ to supersymmetry breaking and extensions that generate an effective $\mu$ parameter dynamically. An effective $\mu$ term can also be generated through modulus stabilization in a modular flavour model~\cite{Fan:2024modulus}.

The next-to-minimal supersymmetric standard model (NMSSM) realizes the second possibility by coupling the two Higgs doublets to a gauge singlet.  In the scale-invariant construction, the singlet vacuum expectation value (VEV) generates an effective $\mu$ parameter and also modifies the Higgs potential at tree level \cite{Ellwanger:2009dp,Maniatis:2009re}.  This setup has been extensively studied through its Higgs and neutralino phenomenology, its interpretation of the Higgs boson near $125$ GeV and its collider signatures \cite{Miller:2003ay,King:2012is,Gunion:2012zd,Carena:2015moc,Ellwanger:2018zxt}. The usual $Z_3$-invariant formulation faces the familiar domain wall and singlet potential issues. Notably, generalized NMSSM constructions include renormalizable singlet terms and explicit Higgs bilinears, providing a broader setting for relating the Higgs mass parameters to singlet dynamics \cite{Abel:1996cr,Panagiotakopoulos:1998yw,Kolda:1998wq,Ross:2011xv,Ross:2012nr,Hollik:2018wrr,Ellwanger:2008py}.   The NMSSM has been studied in deflected anomaly mediation and generalized deflected mirage mediation~\cite{Ning:2017amsb,Du:2017amsb,Du:2018mirage}. Collider and dark matter constraints have also been studied in the semi-constrained NMSSM~\cite{Wang:2018semiconstrained}. Additional Higgs states and electroweakinos in the NMSSM and generalized NMSSM have been discussed in Refs.~\cite{Ellwanger:2024relic,Ellwanger:2024ewinos,Heng:2025light}.

Neutrino oscillation data provide an independent reason to  extend supersymmetric models to account for neutrino masses. NMSSM realizations include $R$-parity violation and higher-dimensional operators \cite{Abada:2006,Gogoladze:2008}, as well as right-handed-neutrino extensions with consequences for neutrino masses, the Higgs spectrum and dark matter \cite{Das:2010,Tang:2014,Cao:2018,Cao:2020}. Inverse seesaw extensions offer further connections to Higgs masses, sneutrino dark matter and leptonic observables \cite{Gogoladze:2012,Cao:2017,Dao:2021,Dao:2022}.

A distinct possibility is the type-II seesaw mechanism, in which integrating out hypercharged electroweak triplets generates the lepton-number-violating Weinberg operator \cite{Weinberg:1979sa,Magg:1980ut,Schechter:1980gr,Lazarides:1980nt,Mohapatra:1980yp,Rossi:2002zb}. Supersymmetric realizations connect neutrino masses to lepton flavour violation and supersymmetry breaking. The triplets can also act as gauge-mediation messengers \cite{Giudice:1998bp,Joaquim:2006uz}. The type-II construction in Ref.~\cite{Li:2020bpp} develops this connection within the NMSSM and provides a theoretical basis for the messenger interpretation considered here.

Neutralino dark matter provides a complementary probe of the electroweakino sector.  Spin-independent dark matter direct detection probes neutralino--Higgs couplings through the Higgs-exchange contribution to neutralino--nucleon scattering.  Singlino-dominated neutralinos can satisfy the direct detection bounds with small scattering cross sections, while annihilation into singlet scalars or coannihilation can reduce their relic density \cite{Cao:2021,Cao:2022,Meng:2024puq,Adhikary:2024nlq}. The relic density and direct detection cross sections of bino- and singlino-dominated neutralino dark matter with a Higgsino component have also been studied in generalized deflected mirage mediation~\cite{Du:2018mirage}. Singlino dark matter and photon signatures at the LHC~\cite{Roy:2024photons}, light bino dark matter in the $Z_3$-invariant NMSSM~\cite{Zhou:2025bino}, and light dark matter in the generalized NMSSM~\cite{Li:2025lightDM} have also been studied. Higgs cascade decays can provide a complementary collider probe of light neutralino dark matter in the semi-constrained NMSSM~\cite{Dong:2026cascade}.

In this work, we choose a singlet basis in which the supersymmetric triplet mass is generated by the  singlet VEV.  The same  VEV contributes to the effective $\mu$ parameter, whose soft-scale value requires a correlated cancellation.  After integrating out the triplets, the low-energy effective field theory (EFT) is a generalized NMSSM supplemented by the type-II Weinberg operator.  We test this construction against theoretical and experimental constraints using bino-like and singlino-like neutralino lightest supersymmetric particles (LSPs) as probes of the low-energy spectrum.   We specify the correlation that fixes $\mu_{\rm eff}$ as an effective boundary input for the low-energy analysis.

The remainder of this paper is organized as follows. Section~\ref{sec:model-eft-construction} introduces the generalized NMSSM and its type-II seesaw sector.  Section~\ref{sec:threshold-consistency} discusses a possible gauge-mediation interpretation of the triplet threshold. The phenomenological analysis is presented in Section~\ref{sec:bino-singlino-dm}. We conclude in Section~\ref{sec:conclusions}.

\section{Generalized NMSSM with a type-II threshold}
\label{sec:model-eft-construction}

\subsection{Correlated singlet scales and the \texorpdfstring{$\mu$}{mu} term}
\label{sec:singlet-bilinear}

We start from the MSSM chiral superfields and add one gauge singlet $S$ together with a vectorlike pair of triplet superfields under $SU(3)_C\times SU(2)_L\times U(1)_Y$,
\begin{equation}
  T\sim(1,3,+1),\qquad \bar T\sim(1,3,-1).
\end{equation}
In terms of fields with definite electric charge, we use
\begin{equation}
 T=\begin{pmatrix}
 T^+/\sqrt2 & -T^{++}\\ T^0 & -T^+/\sqrt2
 \end{pmatrix},\qquad
 \bar T=\begin{pmatrix}
 \bar T^-/\sqrt2 & -\bar T^0\\ \bar T^{--} & -\bar T^-/\sqrt2
 \end{pmatrix}.
 \label{eq:triplet-components}
\end{equation}

The part of the superpotential relevant for the $\mu$ term, the triplet threshold and the type-II seesaw matching is
\begin{align}
  W ={}& W_{\rm Yukawa}
  +\mu_0 H_uH_d+\lambda S H_uH_d
  +L_1S+\frac{M_S}{2}S^2+\frac{\kappa}{3}S^3
  +\left(M_T^{(0)}+\lambda_T S\right)\bar T T               \notag\\
  &+\chi_u H_u\bar T H_u+\chi_d H_d T H_d
  +(\chi_l)_{ij}L_i T L_j ,
  \label{eq:superpotential}
\end{align}
where $W_{\rm Yukawa}$ contains the MSSM Yukawa interactions. We use $\epsilon=i\sigma_2$ and define $A T B=A^T\epsilon T B$, $A\bar T B=A^T\epsilon\bar T B$ and $\bar T T=-\Tr(\bar T T)$. Repeated flavour indices are summed over all $i,j$, and the symmetric matrix $\chi_l^{ij}$ follows a convention that absorbs the customary factor of $1/2$. We work in the charged-lepton mass basis. This convention accounts for the factor of two in the Weinberg coefficient below. 

We impose $R$-parity and assume a neutralino LSP, which is stable within the EFT. All parameters and vacuum expectation values are taken to be real, and we restrict the discussion to the CP-conserving case. Our convention for the vacuum expectation values is
\begin{equation}
  \langle S\rangle=\frac{v_S}{\sqrt2},\quad
  \langle H_u^0\rangle=\frac{v_u}{\sqrt2},\quad
  \langle H_d^0\rangle=\frac{v_d}{\sqrt2}.
\end{equation}
The coefficient $\mu_0$ in Eq.~\eqref{eq:superpotential} denotes the matching-scale boundary value $\mu(M_T)$; its running value at a general scale is $\mu(Q)$. The following basis-invariant supersymmetric mass combinations are defined in terms of running parameters evaluated at a common renormalization scale,
\begin{equation}
  {\mueff(Q)=\mu(Q)+\frac{\lambda(Q)v_S(Q)}{\sqrt2}},
  \qquad
  M_T=M_T^{(0)}+\frac{\lambda_T v_S}{\sqrt2}.
  \label{eq:mueff-mt}
\end{equation}
We choose the triplet phases so that $M_T>0$. This common running supersymmetric mass defines the tree-level matching threshold. Higher-order threshold corrections can shift the corresponding pole masses. Since $S$ is a gauge singlet, the constant translation $S=S'+\delta$ changes the superpotential parameters according to
\begin{equation}
\begin{aligned}
 \mu'&=\mu+\lambda\delta,&
 M_T^{(0)\prime}&=M_T^{(0)}+\lambda_T\delta,\\
 M_S'&=M_S+2\kappa\delta,&
 L_1'&=L_1+M_S\delta+\kappa\delta^2,
\end{aligned}
\qquad
\frac{v_S'}{\sqrt2}=\frac{v_S}{\sqrt2}-\delta .
\label{eq:singlet-shift}
\end{equation}
The corresponding soft coefficients are redefined by the same field translation.  Both masses in Eq.~\eqref{eq:mueff-mt} are invariant.  For $\lambda\lambda_T\ne0$, the combination
\begin{equation}
 {\cal I}_{T\mu}=M_T^{(0)}-\frac{\lambda_T}{\lambda}\mu
 =M_T-\frac{\lambda_T}{\lambda}\mueff
 \label{eq:basis-invariant-mass}
\end{equation}
is also invariant.  We use the matching basis $M_T^{(0)}(M_T)=0$, obtained from Eq.~\eqref{eq:singlet-shift} with $\delta=-M_T^{(0)}/\lambda_T$.  In this basis the triplet threshold is $M_T=\lambda_T(M_T)v_S(M_T)/\sqrt2$.

This hierarchy differs from the usual scale-invariant NMSSM limit. For $|\lambda v_S|/\sqrt2\gg\msusy$, a soft-scale $\mueff$ requires $\mu=-\lambda v_S/\sqrt2+\mathcal O(\msusy)$ at a common scale. We impose this correlation as an effective boundary condition linking the $\mu$ problem to the singlet-generated type-II threshold.

At a fixed renormalization scale, the singlet translation in Eq.~\eqref{eq:singlet-shift} with $\delta=v_S/\sqrt2$ makes $\mu_{\rm eff}$ the explicit coefficient of $H_uH_d$. In the supersymmetric Wilsonian action, the non-renormalization theorem protects the holomorphic superpotential against independent additive perturbative corrections~\cite{Seiberg:1993vc,Martin:1997ns}. Canonically normalized parameters run through wave-function renormalization. The boundary correlation fixes the small bilinear coefficient at the chosen scale. The running combination $\mu_{\rm eff}(Q)$ also depends on $v_S(Q)$, which is determined by the renormalized scalar potential including soft SUSY breaking. Vacuum shifts and finite threshold corrections therefore enter the relation between this combination and the low-energy spectrum.

The soft SUSY-breaking terms retained in the light-field theory are defined by
\begin{align}
-\mathcal L_{\rm soft}\supset{}&
m_{H_u}^2|H_u|^2+m_{H_d}^2|H_d|^2+m_{S,{\rm soft}}^2|S|^2+\sum_{\widetilde f}
\widetilde f^\dagger m_{\widetilde f}^2\widetilde f \notag\\
&+\frac12\left(M_1\widetilde B\widetilde B
+M_2\widetilde W^a\widetilde W^a+M_3\widetilde g^A\widetilde g^A
+\text{h.c.}\right) \notag\\
&+\left[
\begin{aligned}
&B_{\mu_0}H_uH_d+\frac{B_{M_S}}{2}S^2
 +T_\lambda S H_uH_d+\frac{T_\kappa}{3}S^3+t_S S\\
&+\widetilde u\,T_u\widetilde QH_u
 -\widetilde d\,T_d\widetilde QH_d
 -\widetilde e\,T_e\widetilde LH_d+\text{h.c.}
\end{aligned}
\right],
\label{eq:light-soft-terms}
\end{align}
where $\widetilde f$ denotes the sfermion multiplets and $t_S$ is the dimension-three soft singlet tadpole, distinct from the dimension-two superpotential coefficient $L_1$. Equation~\eqref{eq:light-soft-terms} contains the matched light-field soft parameters, including the running tadpole coefficient.

The explicit supersymmetric mass $\mu_0H_uH_d$ breaks the ordinary $Z_3$ symmetry of the scale-invariant NMSSM. The generalized singlet terms allow a large $v_S$ with light singlet-dominated states, subject to the neutral tadpole and scalar mass constraints \cite{Abel:1996cr,Panagiotakopoulos:1998yw,Ross:2012nr}. For a chosen $v_S$, the three neutral tadpole equations determine $m_{H_d}^2$, $m_{H_u}^2$ and $m_{S,{\rm soft}}^2$, while $L_1$ and $t_S$ remain inputs. In addition to stationarity, we require
\begin{equation}
  m_{S,{\rm soft}}^2=\mathcal O(\msusy^2),
  \qquad
  m_{h_S}^2,\;m_{a_S}^2=\mathcal O(\msusy^2),
\end{equation}
at the stationary point. Here $h_S$ and $a_S$ denote singlet-dominated CP-even and CP-odd scalar eigenstates. Their masses and doublet mixing impose conditions independent of the $\mueff$ cancellation.

For $s=v_S/\sqrt2$, the tree-level holomorphic Higgs bilinear in the scalar potential is $B_{\mu,{\rm eff}}=B_{\mu_0}+T_\lambda s+\lambda(L_1+M_Ss+\kappa s^2)$. Electroweak symmetry breaking requires
\begin{equation}
  |\mueff|=\mathcal O(\mew\hbox{--}\msusy),
  \qquad
  B_{\mu,{\rm eff}}=\mathcal O(\msusy^2).
  \label{eq:higgs-conditions}
\end{equation}
The Higgs soft masses $m_{H_u}^2$ and $m_{H_d}^2$ likewise remain of order $\msusy^2$.

\subsection{Triplet matching and neutrino mass}

We perform tree-level matching in the supersymmetric decoupling limit, with $M_T$ well above the retained light-field masses. Expanding $S=v_S/\sqrt2+\widehat S$ gives the triplet bilinear coefficient $M_T+\lambda_T\widehat S$. Suppressing the $SU(2)_L$ indices, the leading equations in $1/M_T$ are
\begin{align}
  \frac{\partial W}{\partial T}
  &=M_T\bar T+\chi_d H_dH_d+\chi_l LL=0,\\
  \frac{\partial W}{\partial \bar T}
  &=M_T T+\chi_u H_uH_u=0.
  \label{F-flat:T}
\end{align}

Substituting the solutions into Eq.~\eqref{eq:superpotential} gives, at leading order in $M_T^{-1}$,
\begin{equation}
  W_{\rm eff}\supset
  -\frac{(\chi_l)_{ij}\chi_u}{M_T}(L_iH_u)(L_jH_u)
  -\frac{\chi_d\chi_u}{M_T}(H_dH_u)^2.
  \label{eq:explicit-triplet-matching}
\end{equation}
We define the operator normalization by
\begin{equation}
  W_{\rm eff}\supset
  \frac{1}{2}C_\nu^{ij}(L_iH_u)(L_jH_u)
  +C_H(H_dH_u)^2,
  \label{eq:weff}
\end{equation}
where $(AB)=A^T\epsilon B$.  With the field and contraction conventions in Eqs.~\eqref{eq:triplet-components} and \eqref{eq:superpotential}, the matching coefficients are
\begin{equation}
  C_\nu^{ij}(M_T)=-\frac{2(\chi_l)_{ij}\chi_u}{M_T},\qquad
  C_H(M_T)=-\frac{\chi_d\chi_u}{M_T}.
  \label{eq:matching-coefficients}
\end{equation}

The factor of two in $C_\nu^{ij}$ follows directly from the operator normalization in Eq.~\eqref{eq:weff}.  Comparing the coefficient of the symmetric product $(L_iH_u)(L_jH_u)$ in Eqs.~\eqref{eq:explicit-triplet-matching} and \eqref{eq:weff} gives $C_\nu^{ij}/2=-(\chi_l)_{ij}\chi_u/M_T$. Below $M_T$ the superpotential becomes
\begin{equation}
  W_< = W_{\rm GNMSSM}
  +\frac{1}{2}C_\nu^{ij}(L_iH_u)(L_jH_u)
  +C_H(H_dH_u)^2+{\cal O}(M_T^{-2}) ,
  \label{eq:low-energy-superpotential}
\end{equation}
where $W_{\rm GNMSSM}$ denotes the generalized NMSSM superpotential containing the light fields.  Terms of order $M_T^{-2}$ include operators with additional insertions of $\widehat S$.  The neutrino mass scale is then
\begin{equation}
  (m_\nu)_{ij}=\frac{1}{2}C_\nu^{ij}v_u^2
  \simeq-\frac{(\chi_l)_{ij}\chi_u v_u^2}{M_T}.
  \label{eq:mnu-eft}
\end{equation}
This is the type-II seesaw contribution in the matched EFT. The Weinberg operator encodes its leading lepton-number-violating effect, with flavour structure determined by the symmetric matrix $\chi_l$. For quantities evaluated at the common scale $Q=M_T$, tree-level inversion gives
\begin{equation}
  \chi_l=-\frac{M_T}{\chi_u v_u^2}\,
  U_\nu^*\operatorname{diag}(m_1,m_2,m_3)U_\nu^\dagger.
  \label{eq:chil-flavour}
\end{equation}
We adopt the PDG convention. Takagi factorization gives $U_\nu^T m_\nu U_\nu=\allowbreak\operatorname{diag}(m_1,m_2,m_3)$ with $m_i\ge0$. The complex conjugations follow from this factorization. CP-conserving flavour inputs and Majorana CP parities are chosen consistently with real $\chi_l$. In the numerical analysis, we evaluate the tree-level relation with the stated low-energy flavour inputs and a fixed electroweak vacuum expectation value, defining the neutrino reconstruction at leading matching order.

The theoretical scale dependence of $C_\nu$ between $M_T$ and the light-supermultiplet thresholds is described, at one loop with GUT-normalized hypercharge, by
\begin{align}
16\pi^2\frac{dC_\nu}{d\ln Q}={}&
(Y_e^\dagger Y_e)^T C_\nu+C_\nu(Y_e^\dagger Y_e)\notag\\
&+\left[6\operatorname{Tr}(Y_u^\dagger Y_u)+2|\lambda|^2
-\frac65g_1^2-6g_2^2\right]C_\nu .
\label{eq:gnmssm-weinberg-rge}
\end{align}
Equation~\eqref{eq:gnmssm-weinberg-rge} combines the MSSM operator evolution with the NMSSM Higgs wave-function contribution \cite{Antusch:2001vn,Ellwanger:2009dp}. The additional $2|\lambda|^2$ term arises from the two $H_u$ fields in the operator. Small $\chi_u$ is technically natural for $\chi_d=T_{\chi_d}=0$ and aligned lepton-number-violating soft trilinears, $T_{\chi_u}=A_{\chi_u}\chi_u$ with finite $A_{\chi_u}$. Here $T_{\chi_u}$ and $T_{\chi_d}$ are the soft coefficients of $H_u\bar T H_u$ and $H_d T H_d$. With $L(T)=-2$ and $L(\bar T)=+2$, the limit $\chi_u\to0$ removes the associated soft source and restores lepton number while $\chi_l$ remains finite.

With $\langle T^0\rangle=v_T/\sqrt2$ and $\langle\bar T^0\rangle=v_{\bar T}/\sqrt2$, the leading supersymmetric decoupling solution of Eq.~\eqref{F-flat:T}, at a vacuum with vanishing slepton  VEVs, gives
\begin{align}
 v_T^{\rm ind}&\simeq-\frac{\chi_u v_u^2}{\sqrt2M_T},&
 v_{\bar T}^{\rm ind}&\simeq-\frac{\chi_d v_d^2}{\sqrt2M_T}.
 \label{eq:induced-triplet-vevs}
\end{align}
These expressions give the leading terms in the heavy-mass expansion; electroweak and soft-breaking effects shift the triplet VEVs. In the aligned $\chi_d=T_{\chi_d}=0$ limit used below, the displayed leading term in $v_{\bar T}$ vanishes, while subleading induced contributions can remain. Corrections to the type-II relation are subleading when $|\mueff|$, the triplet soft mass scales and $|A_{\chi_u}|$ are small compared with $M_T$, and the holomorphic scalar splitting is small compared with $M_T^2$. The induced $v_T$ is the full-theory representation of the Weinberg-operator contribution in Eq.~\eqref{eq:mnu-eft}.

 A nonzero VEV of a complex triplet with hypercharge one breaks custodial symmetry and modifies the rho parameter at tree level.  With $v^2=v_u^2+v_d^2$ and the triplet VEV convention used above,
\begin{equation}
  \rho_{\rm tree}=
  \frac{v^2+2(v_T^2+v_{\bar T}^2)}
       {v^2+4(v_T^2+v_{\bar T}^2)}.
\end{equation}
For small induced values, the deviation is
\begin{equation}
  \rho-1 \simeq
  -\frac{2(v_T^2+v_{\bar T}^2)}{v^2}.
  \label{eq:rho}
\end{equation}
Electroweak precision data require $v_T$ and $v_{\bar T}$ to be well below the electroweak scale. At leading order, neutrino masses constrain the matrix product $m_\nu=\sqrt2\,v_T\chi_l$ through
\begin{equation}
  m_k=\sqrt2\,|v_T|\,\sigma_k(\chi_l),\qquad k=1,2,3,
  \label{eq:nu-vt-bound}
\end{equation}
where $\sigma_k(\chi_l)$ are the singular values ordered consistently with the nonnegative Majorana masses at the same scale.  The neutrino-mass constraint on $v_T$ is set by $\chi_l$, while $v_{\bar T}$ enters the electroweak-precision constraint.

\section{A possible gauge-mediation interpretation of the triplet threshold}
\label{sec:threshold-consistency}

The vectorlike triplet pair can form part of a gauge-mediation messenger threshold in an ultraviolet completion.  To exhibit this parametric interpretation, introduce a spurion with a vanishing scalar component,
\begin{equation}
  X=\theta^2 F_X,
\end{equation}
and allow the triplet bilinear to couple to both $S$ and $X$,
\begin{equation}
  W\supset\left(M_T^{(0)}+\lambda_T S+k_X X\right)\bar T T .
  \label{eq:messenger-superpotential}
\end{equation}
The supersymmetric messenger mass parameter and its holomorphic SUSY-breaking insertion are
\begin{equation}
  M_{\rm mess}=M_T^{(0)}+\lambda_T\langle S\rangle=M_T,
  \qquad
  F_{\rm mess}=\lambda_T F_S+k_X F_X .
  \label{eq:messenger-mf}
\end{equation}

Here $M_{\rm mess}$ sets the decoupling scale, while $F_{\rm mess}$ gives the holomorphic insertion with mass dimension two that splits the two messenger scalar eigenstates,
\begin{equation}
  m_\pm^2=|M_{\rm mess}|^2\pm|F_{\rm mess}|
  +\mathcal O(\msusy^2).
  \label{eq:messenger-splitting}
\end{equation}
Neglecting the additional soft mass corrections in Eq.~\eqref{eq:messenger-splitting}, positive messenger scalar masses require
\begin{equation}
  x\equiv\frac{|F_{\rm mess}|}{|M_{\rm mess}|^2}<1.
  \label{eq:messenger-stability}
\end{equation}
The overlap with the supersymmetric matching in Section~\ref{sec:model-eft-construction} additionally requires $x\ll1$ and both triplet scalar masses well above the retained light-field scales.

The leading gauge contributions to the gaugino masses and scalar soft mass-squared parameters at $M_{\rm mess}$ are
\begin{align}
  M_a&=\frac{\alpha_a}{4\pi}\,n_a\,\Lambda_{\rm GM}\,g(x),
  \label{eq:gmsb-gaugino}\\
  m_i^2&=2|\Lambda_{\rm GM}|^2
  \sum_a C_a(i)\left(\frac{\alpha_a}{4\pi}\right)^2n_a f(x),
  \label{eq:gmsb-scalar}
\end{align}
where
\begin{equation}
  \Lambda_{\rm GM}=\frac{F_{\rm mess}}{M_{\rm mess}},
  \qquad g(0)=f(0)=1 .
  \label{eq:lambda-gm}
\end{equation}
The functions $g$ and $f$ are the finite-splitting threshold functions of gauge mediation \cite{Giudice:1998bp,Martin:1996zb}.  Using GUT-normalized hypercharge, one $T+\bar T$ pair with $Y=\pm1$ has messenger indices
\begin{equation}
  n_a=\Delta b_a=\left(\frac{18}{5},\,4,\,0\right).
  \label{eq:triplet-messenger-indices}
\end{equation}

An isolated triplet pair consequently generates bino and wino masses through the leading gauge contributions, while a leading gluino mass requires colored messenger states. The nonzero hypercharge messenger index also gives right-handed squarks a $U(1)_Y$-mediated soft mass through Eq.~\eqref{eq:gmsb-scalar}; colored messengers supply the leading $SU(3)_C$ contributions. 

A possible unified embedding places the triplets in a $\mathbf{15}+\overline{\mathbf{15}}$ pair of $SU(5)$, as in the type-II gauge-mediation construction of Ref.~\cite{Li:2020bpp}. Under $SU(3)_C\times SU(2)_L\times U(1)_Y$,
\begin{align}
 \mathbf{15}&=(\mathbf{6},\mathbf{1},-2/3)\oplus(\mathbf{1},\mathbf{3},+1)\oplus(\mathbf{3},\mathbf{2},1/6),\notag\\
 \overline{\mathbf{15}}&=(\overline{\mathbf{6}},\mathbf{1},2/3)\oplus(\mathbf{1},\mathbf{3},-1)\oplus(\overline{\mathbf{3}},\mathbf{2},-1/6).
 \label{eq:su5-messenger-decomposition}
\end{align}
Messenger--matter interactions can also generate Yukawa-deflected soft terms beyond the minimal gauge contributions in Eqs.~\eqref{eq:gmsb-gaugino} and \eqref{eq:gmsb-scalar} \cite{Chacko:2001km,Li:2020bpp}. Such terms can modify the trilinear couplings and scalar spectrum relevant for the Higgs mass. Their form depends on the ultraviolet messenger couplings.

A TeV-scale soft spectrum corresponds parametrically to $\Lambda_{\rm GM}\sim10^5$ GeV. Together with Eq.~\eqref{eq:messenger-stability}, this gives $M_T>|\Lambda_{\rm GM}|$.  

In the messenger interpretation, $F_{\rm mess}$, the total supersymmetry-breaking scale, the colored messenger content and the correlated soft boundary conditions are ultraviolet inputs. The neutralino dark-matter interpretation applies when the neutralino is the LSP, including the gravitino and hidden-sector states. We adopt independent soft matching inputs at $M_T$ for the phenomenological analysis.

\section{Numerical analysis}
\label{sec:bino-singlino-dm}

\subsection{Numerical setup and constraints}
\label{sec:scan-constraints}

We study the low energy EFT without specifying its UV origin. The following universal boundary conditions at $M_T$ scale are adopted for the soft scalar masses 
\begin{equation}
 m_{\widetilde Q}^2=m_{\widetilde U}^2=m_{\widetilde D}^2
 =m_{\widetilde L}^2=m_{\widetilde E}^2=m_0^2\mathbf{1}~,
\quad \quad
 \label{eq:universal-soft-boundary}
\end{equation}
and universal trilinear coupling $T_f=A_0Y_f$ for $f=u,d,e$. The gaugino masses are set as free parameters. Soft SUSY breaking spectrum of this type can, for example, be generated by scherk-Schwarz mechanism in orbifold GUT model with proper non-trivial boundary conditions and twisting~\cite{Barbieri:2001dm,Han:2013poa}, assuming the compactification scale lies slightly upon the $M_T$ scale. Besides, the concrete values for the bare $\mu_0$ and $B\mu_0$ terms are also predicted by the scherk-Schwarz model. Such a setup of this model is phenomenologically attractive, as it highlight the key phenomenological issues of major interest, including the $\mu$-problem, neutrino mass generation, and the dark matter.


For the singlet--triplet sector, we define $r_\lambda=\lambda^{\rm in}/\lambda_T^{\rm in}$ and take $M_T$, $\lambda_T^{\rm in}$ and $r_\lambda$ as independent inputs. The input parameters are taken in the ranges
\begin{equation}
\begin{gathered}
10^7\leq M_T/{\rm GeV}\leq10^{15},\qquad
{0 < \lambda_T^{\rm in}\leq0.75},\qquad
{0\leq r_\lambda\leq1},\\
0.7\leq m_0/{\rm TeV}\leq15,\qquad
{1\leq\tan\beta\leq50} .
\end{gathered}
\label{eq:common-scan-ranges}
\end{equation}
At the supersymmetry scale $Q_{\rm SUSY}$, the parameters $v_S^{\rm in}=\sqrt2 M_T/\lambda_T^{\rm in}=v_S(Q_{\rm SUSY})$ and $\lambda^{\rm in}=r_\lambda\lambda_T^{\rm in}$ are derived from these inputs, with the additional condition $10^8\leq v_S^{\rm in}/{\rm GeV}\leq10^{16}$. In the full theory, the singlet-generated triplet mass is expressed in terms of parameters evaluated at the common scale $M_T$. Relating the two descriptions requires consistent matching across scales. We set
\begin{equation}
 \kappa=T_\kappa=\chi_d=0 .
 \label{eq:singlet-scan-ranges}
\end{equation}
 For the matter-sector coefficient defined by $T_f=A_0Y_f$, we take $-1.5\leq A_0/m_0\leq1$. The linear singlet boundary parameter satisfies $L_1=-M_Sv_S^{\rm in}/\sqrt2$, together with the electroweak-symmetry-breaking conditions.

We calculate the spectrum, Higgs observables and thermal relic density using the tools of Refs.~\cite{Staub:2013tta,Porod:2003um,Porod:2011nf,HiggsBounds4,HiggsSignals3,Alguero:2023zol}. For the neutrino sector, we use the tree-level type-II relation and the separate three-flavour reconstruction described above.

We choose the correlated boundary value $\mu_0$ to obtain an effective Higgs bilinear of order the soft SUSY masses. At the supersymmetry scale,
\begin{equation}
 \mu_{\rm eff}(Q_{\rm SUSY})=\mu(Q_{\rm SUSY})
 +\frac{\lambda(Q_{\rm SUSY})v_S(Q_{\rm SUSY})}{\sqrt2}.
 \label{eq:mu-eff-susy-scale}
\end{equation}
We classify bino-like and singlino-like LSPs by $|N_{11}|^2\geq0.50$ and $|N_{15}|^2\geq0.50$, respectively, in the basis $(\widetilde B,\widetilde W^0,\widetilde H_d^0,\widetilde H_u^0,\widetilde S)$. The two classes also contain mixed states.

\paragraph{Experimental constraints and flavour matching.}

We require the mass of the SM-like Higgs boson to lie in the range $122$--$128$ GeV. All spectra displayed in the figures satisfy the Higgs-search limits at $95\%$ confidence level from HiggsBounds~5.10.2 and the HiggsSignals~2.6.2 condition $p_{\rm HS}>0.05$. Both constraints are evaluated using the same spectrum.

We impose the sparticle mass requirements $m_{\chi_1^\pm}>103.5$ GeV, $m_{\widetilde g}>2400$ GeV, $m_{\widetilde t_1},m_{\widetilde b_1}>1250$ GeV, first- and second-generation squarks above $1400$ GeV, and $m_{\widetilde\tau_1}>480$ GeV \cite{LEP,ATLAS:2020qlk,ATLAS:2023multib,ATLAS:2020dsf,ATLAS:2021sbottom,ATLAS:2024fub}. These mass thresholds serve as approximate collider preselection criteria, whose interpretation depends on the production and decay topology. The displayed spectra are subject to the conditions listed here. For the benchmark spectra, we supplement these conditions with leading-order reinterpretations of selected LHC searches using CheckMATE~\cite{Dercks:2016npn} and conservative inclusive-production bounds.

We impose the following flavour constraints:
\begin{equation}
 \begin{aligned}
 3.11\times10^{-4}&\leq{\rm BR}(B\to X_s\gamma)\leq3.87\times10^{-4},\\
 2.80\times10^{-9}&\leq{\rm BR}(B_s\to\mu^+\mu^-)\leq3.88\times10^{-9}.
 \end{aligned}
\end{equation}
The dark matter relic density is required to satisfy $0<\Omega_\chi h^2\leq0.144$. Here we require the full dark matter abundance with $0.096\leq\Omega_\chi h^2\leq0.144$ as a reference range~\cite{Planck:2018vyg}.

We require the abundance-weighted spin-independent (SI), pure-proton spin-dependent (SD) and pure-neutron SD cross sections to satisfy the corresponding LZ upper limits separately \cite{LZ:2024}. The direct-detection figures show the abundance-weighted SI cross section together with the official LZ observed SI nucleon reference curve~\cite{LZ:HEPData}, under the equal-proton-and-neutron-coupling assumption. Approximate XENONnT 2023 and PandaX-4T curves are shown for comparison \cite{XENONnT:2023,PandaX:2024}. The separate LZ conditions are used for the direct-detection selection. The more recent XENONnT result is given in Ref.~\cite{XENONnT:2025}. The separate-operator limits constrain the corresponding nucleon couplings, while a general mixed-nucleon response requires a detector-level likelihood.

For the leading three-flavour matching, we assume normal ordering with $m_1=0.010$ eV and the CP-conserving phase $\delta_{\rm CP}=\pi$. The NuFIT~6.0 IC24 inputs, including Super-Kamiokande atmospheric data, are $\Delta m_{21}^2=7.49\times10^{-5}$ eV$^2$, $\Delta m_{31}^2=2.513\times10^{-3}$ eV$^2$ and $(\sin^2\theta_{12},\sin^2\theta_{13},\sin^2\theta_{23})=(0.308,0.02215,0.470)$ \cite{NuFIT:2024}. We choose the lightest mass and phase for this scenario. With $v=246.21965$ GeV and $v_u=v\sin\beta$, the leading matching relation is
\begin{equation}
 \chi_l=-\frac{M_T}{\chi_u v_u^2}
 U_\nu^*\operatorname{diag}(m_1,m_2,m_3)U_\nu^\dagger,\qquad
 \max(\|\chi_l\|_2,|\chi_u|)<\sqrt{4\pi}.
 \label{eq:three-flavour-screen}
\end{equation}
Here, the norm denotes the largest singular value. The resulting mass sum is about $0.0743$ eV. For flavor-diagonal soft matching at $M_T$, we compare the leading LFV estimates with present bounds, including $\mathrm{BR}(\mu\to e\gamma)<1.5\times10^{-13}$ \cite{MEGII:2025}. Additional flavor-violating soft terms depend on the ultraviolet completion.

The results below refer to the spectrum and experimental conditions stated above. Electroweak-vacuum stability is a separate condition on the scalar potential in a consistent renormalization scheme.

\subsection{Mass spectrum and dark matter}
\label{sec:numerical-two-scales}

The type-II threshold and the neutralino mass depend on different combinations of parameters. With $m_\chi=|m_{\chi^0_1}|$ denoting the LSP mass, the studied bino-like solutions span $m_\chi\simeq95.6$ GeV--$3.25$ TeV and the singlino-like solutions $125$ GeV--$3.31$ TeV. Their triplet input scales reach $9.97\times10^{14}$ and $9.26\times10^{14}$ GeV, respectively. Within the reference relic density range, the largest triplet scales are $4.39\times10^{14}$ and $9.26\times10^{14}$ GeV. This separation of scales follows from the independent electroweakino masses and the correlated Higgs bilinear.

\begin{figure}[tbp]
 \centering
 \includegraphics[width=0.96\textwidth]{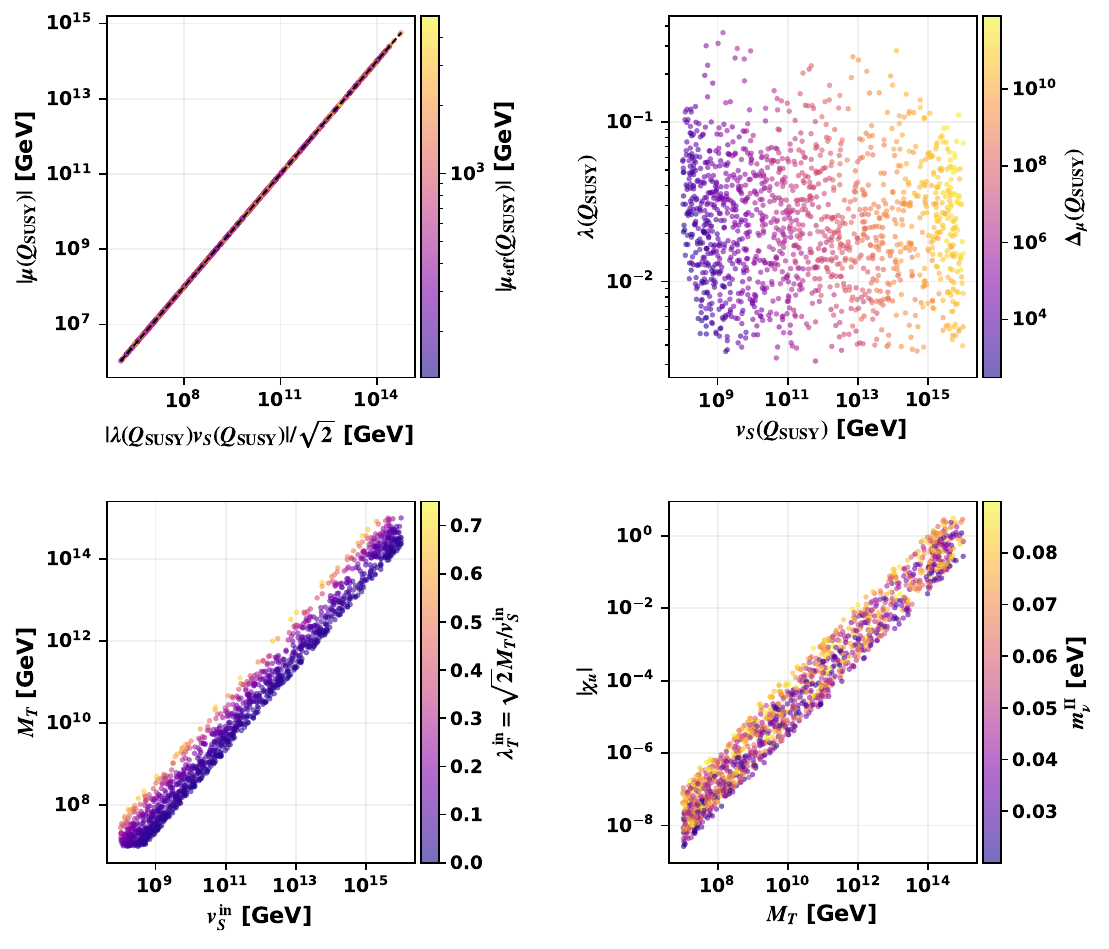}
 \caption{Singlet-sector relations at $Q_{\rm SUSY}$ and the type-II input scale. The upper panels compare $|\mu|$ with $|\lambda v_S|/\sqrt2$ and $v_S$ with $\lambda$; colours denote $|\mu_{\rm eff}|$ and $\Delta_\mu$, respectively. The dashed diagonal corresponds to equal Higgs-bilinear contributions. The lower panels display $M_T$ against $v_S^{\rm in}$, with colour denoting $\lambda_T^{\rm in}$, and $|\chi_u|$ against $M_T$, with colour denoting the leading type-II neutrino mass.}
 \label{fig:mu-typeII-consistency}
\end{figure}

The singlet-sector relations in Fig.~\ref{fig:mu-typeII-consistency} illustrate the correlated Higgs bilinear. The upper panels use running parameters at $Q_{\rm SUSY}$, where the explicit Higgs bilinear nearly cancels the singlet contribution in $\mu_{\rm eff}$. The lower-left panel gives the input relation $M_T=\lambda_T^{\rm in}v_S^{\rm in}/\sqrt2$, with $v_S^{\rm in}=v_S(Q_{\rm SUSY})$ as specified above.

We characterize the cancellation in the upper-left panel at the common scale by
\begin{equation}
 \Delta_\mu(Q_{\rm SUSY})=
 \frac{\max\!\left(|\mu(Q_{\rm SUSY})|,
 |\lambda(Q_{\rm SUSY})v_S(Q_{\rm SUSY})/\sqrt2|\right)}
 {|\mu_{\rm eff}(Q_{\rm SUSY})|}.
 \label{eq:mu-cancellation-diagnostic}
\end{equation}
For the studied bino-like and singlino-like solutions, the median values are $1.23\times10^6$ and $4.96\times10^6$, respectively, showing a substantial cancellation. This basis-dependent ratio characterizes the cancellation for the chosen boundary inputs. Parameter sensitivities provide a separate measure of naturalness. The Higgs-bilinear correlation is specified through the boundary inputs. The supersymmetric renormalization structure and the conditions on the soft-breaking vacuum are discussed in Section~\ref{sec:singlet-bilinear}.

\begin{figure}[tbp]
 \centering
 \includegraphics[width=0.96\textwidth]{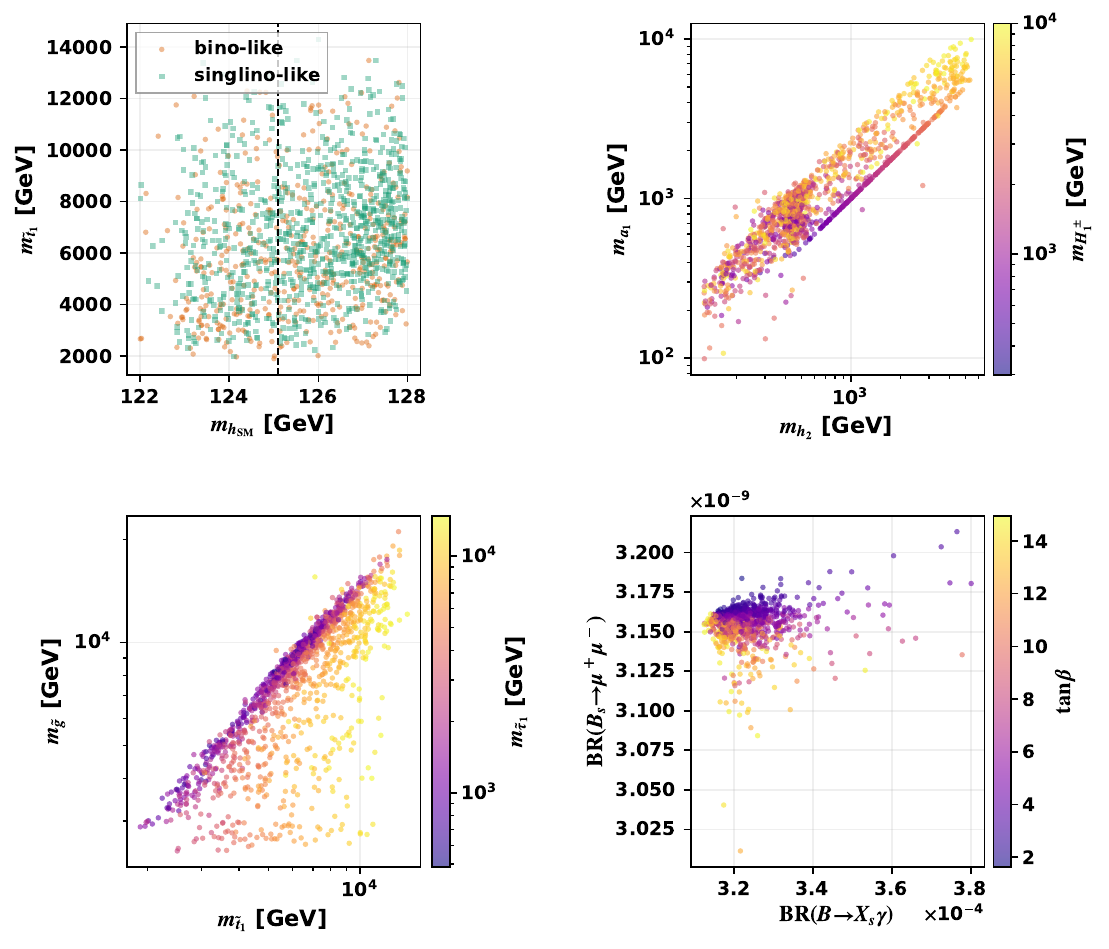}
 \caption{Higgs and coloured-sparticle masses and flavour observables for both LSP classes. The panels compare the SM-like Higgs and lighter stop masses, the $h_2$ and $a_1$ masses, the lighter stop and gluino masses, and ${\rm BR}(B\to X_s\gamma)$ and ${\rm BR}(B_s\to\mu^+\mu^-)$. The upper-left legend labels the LSP classes, and the dashed line marks $m_h=125.09$ GeV. Colours in the remaining panels denote the charged-Higgs mass, lighter-stau mass and $\tan\beta$, respectively.}
 \label{fig:shared-spectrum}
\end{figure}

The Higgs and sparticle mass spectra and flavour observables for both LSP classes are shown in Fig.~\ref{fig:shared-spectrum}. The lighter stop mass spans $1.89$--$13.5$ TeV for the studied bino-like solutions and $1.98$--$14.3$ TeV for the singlino-like solutions. The lighter additional CP-even state $h_2$ spans $0.148$--$5.41$ TeV and $0.128$--$2.72$ TeV, respectively, and $a_1$ spans $0.218$--$9.93$ TeV and $0.0993$--$3.90$ TeV. These extrema can belong to different solutions. The corresponding ranges of $\mathrm{BR}(B\to X_s\gamma)$ are $(3.12$--$3.76)\times10^{-4}$ and $(3.12$--$3.80)\times10^{-4}$, and those of $\mathrm{BR}(B_s\to\mu^+\mu^-)$ are $(3.01$--$3.21)\times10^{-9}$ and $(3.04$--$3.20)\times10^{-9}$. Heavy coloured superpartners can therefore coexist with the Higgs-mediated dark matter mechanisms discussed below under the adopted flavour conditions.

\paragraph{Bino-like neutralino dark matter.}
\label{sec:bino-results}

A bino interacts with Standard Model fermions through sfermion exchange. A Higgsino component also allows scattering through Higgs exchange and opens additional annihilation channels. Efficient thermal depletion and the direct detection bounds therefore constrain different combinations of masses and mixings. We denote the bino, singlino and total Higgsino fractions by $f_B=|N_{11}|^2$, $f_S=|N_{15}|^2$ and $f_H=|N_{13}|^2+|N_{14}|^2$, respectively. Within the reference relic density range, the studied bino-like solutions satisfy $f_B\geq0.9938$.

\begin{figure}[tbp]
 \centering
 \includegraphics[width=0.96\textwidth]{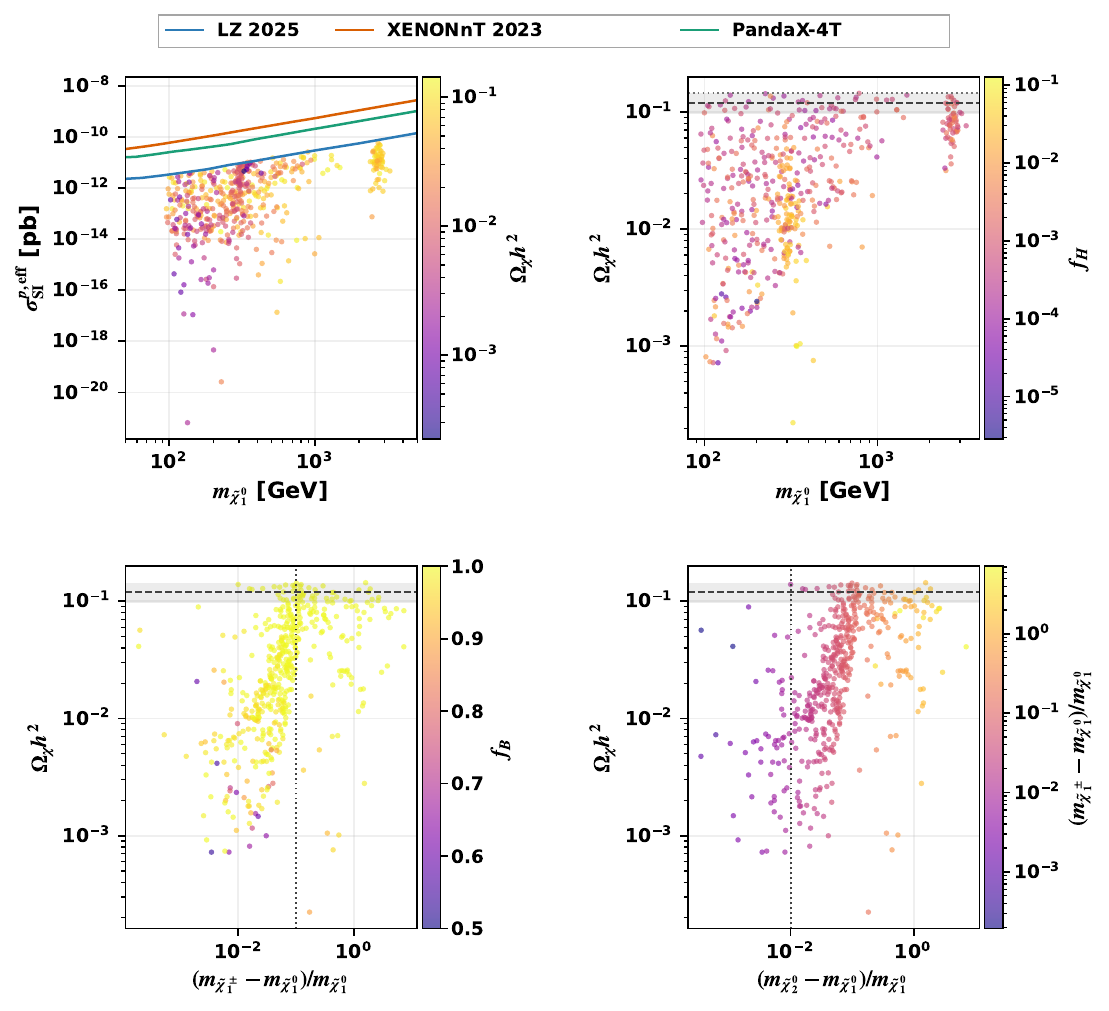}
 \caption{Bino-like dark matter. The upper panels show the abundance-weighted SI cross section and relic density versus the LSP mass; the lower panels show the relic density versus the relative mass gaps of the lightest chargino and second neutralino. The SI curves give the LZ result and approximate XENONnT 2023 and PandaX-4T results \cite{LZ:2024,XENONnT:2023,PandaX:2024}. The relic density band and central line correspond to $0.096\leq\Omega_\chi h^2\leq0.144$ and $\Omega_\chi h^2=0.120$. Colours denote $\Omega_\chi h^2$, $f_H$, $f_B$ and the chargino relative gap, respectively. The selection uses SI and pure-nucleon SD limits.}
 \label{fig:bino-mechanisms}
\end{figure}

Using absolute physical masses, the two relative gaps in Fig.~\ref{fig:bino-mechanisms} are $(|m_{\chi_1^\pm}|-m_\chi)/m_\chi$ and $(|m_{\chi_2^0}|-m_\chi)/m_\chi$. A small gap reduces the Boltzmann suppression of the heavier state during freeze-out. The relic density also depends on the partner interactions and on the conversion reactions that maintain chemical equilibrium.

To compare the two neutralino classes, we define the ratio of the effective SI cross section to the adopted mass-dependent LZ reference curve in Fig.~\ref{fig:bino-mechanisms}, which is also used for the benchmark ratios, by
\begin{equation}
 R_{\rm SI}=\frac{\sigma^{p,\mathrm{eff}}_{\rm SI}}{\sigma_{\rm SI}^{\rm ref}(m_\chi)}.
 \label{eq:si-ratio}
\end{equation}

\paragraph{Singlino-like neutralino dark matter.}
\label{sec:singlino-results}

\begin{figure}[tbp]
 \centering
 \includegraphics[width=0.96\textwidth]{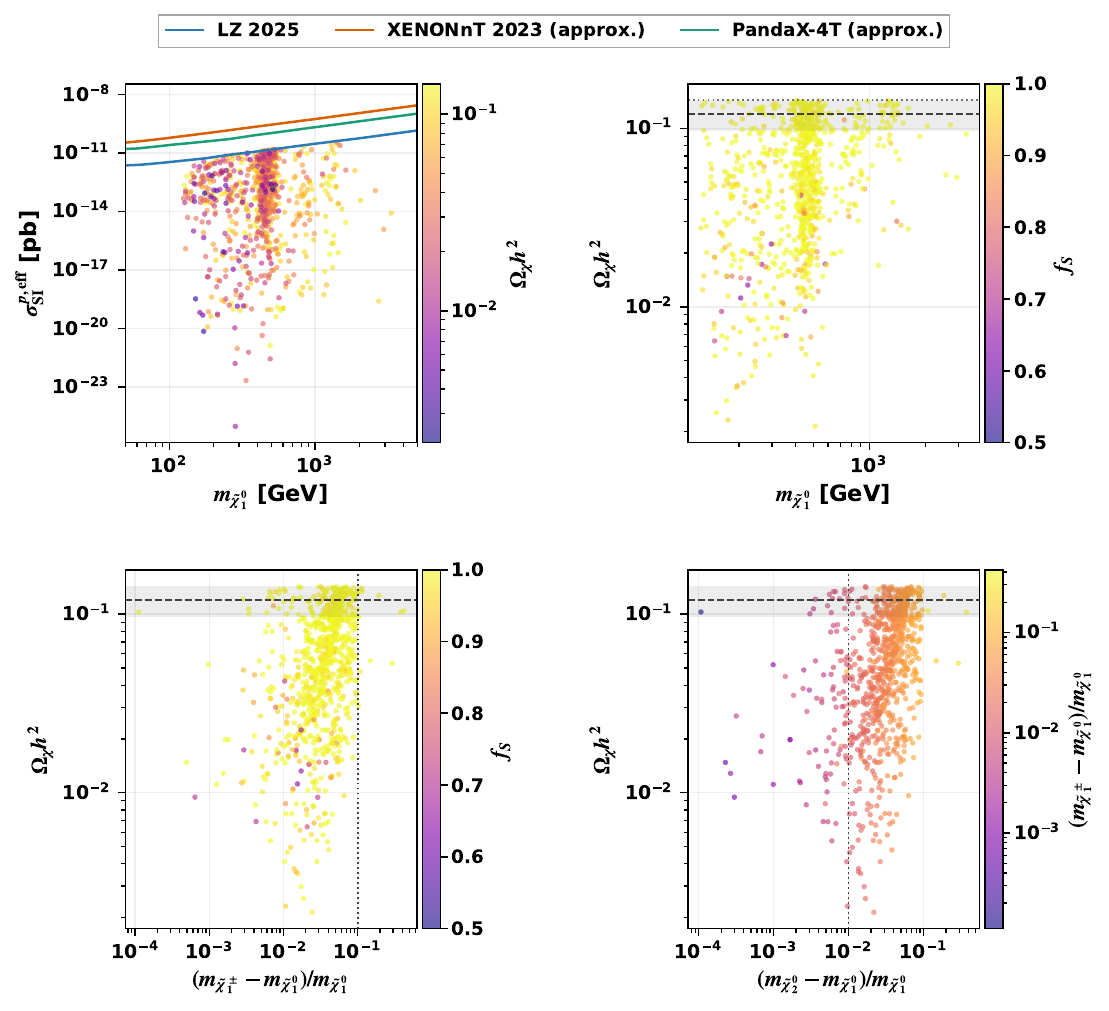}
 \caption{Singlino-like dark matter. The upper panels show the abundance-weighted SI cross section and relic density versus the LSP mass; the lower panels show the relic density versus the relative mass gaps of the lightest chargino and second neutralino. The reference curves and relic density band are the same as in Fig.~\ref{fig:bino-mechanisms}. Colours denote $\Omega_\chi h^2$ in the upper-left panel, $f_S$ in the upper-right and lower-left panels, and the chargino relative gap in the lower-right panel.}
 \label{fig:singlino-mechanisms}
\end{figure}

For $\kappa=0$, $M_S$ primarily controls the singlino diagonal entry in the neutralino mass matrix. Higgsino and singlet--doublet Higgs mixing determine its couplings for scattering on nucleons. The SI cross sections, relic densities and electroweakino mass gaps are shown in Fig.~\ref{fig:singlino-mechanisms}. Within the reference relic density range, the studied solutions satisfy $f_S\geq0.8824$. Coannihilation with nearby supersymmetric particles can reduce the relic density while the scattering cross section remains small.

\subsection{Annihilation and coannihilation}

Neutralino annihilation can be enhanced when the centre-of-mass energy approaches the mass of an $s$-channel Higgs mediator. The enhancement depends on the masses, widths, couplings and thermal distribution \cite{Griest:1990kh}. As discussed below, BP-B0 and BP-S0 illustrate spectra consistent with a Higgs funnel for bino-like and singlino-like neutralinos, respectively. Coannihilation is classified by its incoming particles, and the corresponding reactions can also be resonant.

\begin{figure}[tbp]
 \centering
 \includegraphics[width=0.96\textwidth]{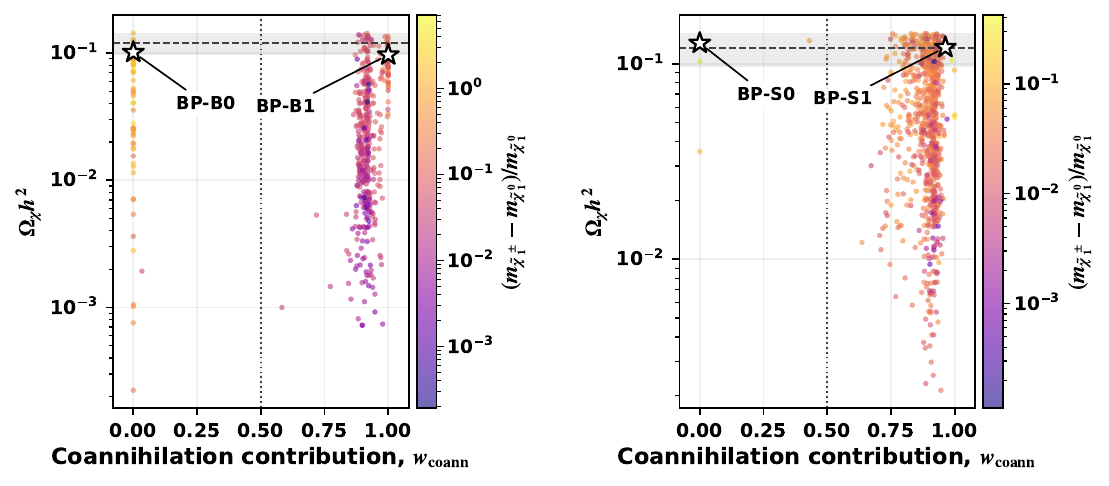}
 \caption{Relic density as a function of $w_{\rm coann}$ for bino-like (left) and singlino-like (right) dark matter. Colour denotes the relative lightest-chargino mass gap, and labelled markers identify the benchmarks. The shaded band and central line are the same as in Fig.~\ref{fig:bino-mechanisms}. The coannihilation contribution sums the leading inverse-relic-density channel weights under chemical equilibrium.}
 \label{fig:thermal-channels}
\end{figure}
\FloatBarrier

We define the coannihilation contribution $w_{\rm coann}$ as the sum of the inverse-relic-density weights of the leading thermal channels with at least one incoming supersymmetric particle heavier than the LSP. The heavier partners can be electroweakinos, gluinos or sfermions. A large $w_{\rm coann}$ indicates that these partner reactions dominate the listed thermal channels. Its relation to the relic density and lightest-chargino mass gap is shown for both LSP classes in Fig.~\ref{fig:thermal-channels}. For the solutions near $w_{\rm coann}=0$ and $1$, the listed thermal reactions are dominated by LSP pairs and by heavier partners, respectively. The coannihilation contribution classifies the initial states; identifying a Higgs funnel requires the mediator masses, widths and neutralino couplings. The benchmarks below present these ingredients explicitly. The standard coannihilation treatment assumes chemical equilibrium during freeze-out \cite{Griest:1990kh,Edsjo:1997bg}. This requires conversion reactions to be faster than the Hubble expansion, $\Gamma_{\rm conv}\gg H$. The condition depends on both the conversion interactions and the mass splittings, particularly in weakly coupled scenarios \cite{Garny:2017rxs,Ellis:2015vaa}.

We define the relative mass gap to the nearest electroweakino by
\begin{equation}
 \delta_{\rm EWino}=
 \min_{j\in\{\chi^0_2,\,\chi^\pm_1\},\,|m_j|>m_\chi}
 \frac{|m_j|-m_\chi}{m_\chi}.
 \label{eq:ewino-gap}
\end{equation}
The minimization runs over the indicated electroweakinos. For the benchmarks we also define $\delta_{\widetilde g}=(|m_{\widetilde g}|-m_\chi)/m_\chi$. A small electroweakino gap can favor partner reactions, as illustrated by BP-S1. Other superpartners can also contribute: BP-B1 illustrates gluino-pair annihilation with a larger electroweakino gap.
\FloatBarrier

\subsection{Benchmark points}
Tables~\ref{tab:benchmark-bino} and \ref{tab:benchmark-singlino} compare LSP-pair annihilation with reactions involving heavier supersymmetric partners. The weights $w_{\chi\chi}$ and $w_{\rm coann}$ sum the listed leading contributions to the inverse relic density; further subleading reactions can also contribute. These weights classify initial states, and the resonance entries specify the relevant Higgs states. Higgs masses appear in the spectrum block, while the widths and threshold offsets follow the state ordering in the resonance block. A dash denotes an unspecified resonance assignment.

Boundary inputs are specified at $M_T$, {while} $\lambda$, $v_S$ and $\mu_{\rm eff}$ are evaluated at $Q_{\rm SUSY}$. The quantity $\lambda_T^{\rm in}$ is an input parameter. The one-flavour input $\chi_l^{\rm in}$ is distinct from the reconstructed three-flavour norm $\|\chi_l\|_2$. Pole masses are positive, $\mu_{\rm eff}$ retains its sign, and $f_H$ includes both Higgsinos. The effective cross sections include relative relic density weighting; $R_{\rm SI}$ uses the official LZ observed nucleon projection. Reaction weights include charge-conjugate channels where appropriate. The symbol $q$ sums over six quark flavours, and $d_i\bar u_i$ sums over three families.

\textbf{Bino-like spectra.} BP-B0 is consistent with Higgs-funnel annihilation in the region of the nearly degenerate $h_2/a_1$ poles. Its $634.5$ GeV LSP has bino and Higgsino fractions of $0.999392$ and $6.076\times10^{-4}$, respectively, with a negligible wino component. The poles $m_{h_2}=1270.64$ GeV and $m_{a_1}=1269.45$ GeV lie within one width of $2m_\chi=1268.99$ GeV; their widths are about $5.74$ and $5.88$ GeV, respectively. All listed channels are LSP-pair annihilations; the $t\bar t$ and $b\bar b$ channels have weights $0.9743$ and $0.01992$, respectively. The pole positions, widths and leading channels support a Higgs-funnel interpretation involving the nearby $h_2$ and $a_1$ states.

BP-B1 has a $3.065$ TeV LSP and a $3.094$ TeV gluino. Despite the larger electroweakino gap, gluino pairs dominate the listed reactions, with weights $0.6316$ for quark pairs and $0.3684$ for gluons. Through the conversion processes discussed above, gluino-pair annihilation can deplete the total relic density \cite{Profumo:2004at}. Sommerfeld enhancement and bound-state formation may modify the coloured-partner contribution \cite{Ellis:2015vaa}.

\textbf{Singlino-like spectra.} BP-S0 is consistent with a thermally accessible $h_3/a_2$ heavy-Higgs funnel for a $385.1$ GeV LSP with singlino fraction $0.9951$. Its Higgsino fraction $0.004241$ and $\lambda=0.0565$ allow a neutralino--Higgs coupling. The poles $m_{h_3}=776.49$ GeV and $m_{a_2}=775.67$ GeV lie $6.19$ and $5.38$ GeV above $2m_\chi=770.30$ GeV, respectively. These offsets exceed the vacuum widths. At $x_f=m_\chi/T_f=24.3$, with freeze-out temperature $T_f\simeq15.8$ GeV, thermal kinetic energy makes the poles accessible~\cite{Griest:1990kh}. The leading LSP-pair weights are $0.7151$ for $t\bar t$, $0.2257$ for $b\bar b$ and $0.04008$ for $\tau^+\tau^-$. The listed coannihilation weight is zero. The $h_3$ and $a_2$ pole positions and the leading channels support this thermally accessible heavy-Higgs-funnel interpretation.

BP-S1 illustrates coannihilation-dominated freeze-out. Its $1436.5$ GeV LSP has singlino fraction $0.9997815$. The predominantly wino second neutralino and the lighter chargino have masses of $1441.7$ and $1440.7$ GeV, with relative gaps of $0.36\%$ and $0.29\%$. The listed partner weight is $0.963331$, and all listed leading channels involve heavier partners. The largest channel groups are $\widetilde\chi^0_2\widetilde\chi^-_1\to d_i\bar u_i$ and $\widetilde\chi^-_1\widetilde\chi^+_1\to q\bar q$, with weights $0.3344$ and $0.1832$. Other reactions of the wino-like partners into gauge bosons also contribute. The relic density is $\Omega_\chi h^2=0.1213$.

For each benchmark, the tables give the HiggsSignals fit $p$-value $p_{\rm HS}$ and the observed ratio $R_{\rm HB}$ for the most-sensitive Higgs analysis. All benchmarks satisfy the adopted HiggsBounds and HiggsSignals criteria evaluated with the same spectra. The tables also give the flavour observables.

\textbf{LHC searches.} For the monojet, jets plus missing transverse momentum, soft-lepton and multilepton searches considered here \cite{ATLAS:2021monojet,ATLAS:2020qlk,ATLAS:2019compressed,ATLAS:2019dilepton,ATLAS:2021trilepton}, the leading-order reinterpretation gives conservative signal-yield bounds for BP-B0 and BP-S0 below the experimental limits. For BP-B1, the conservative inclusive production cross section gives a signal-yield bound below the limits of these searches. The bound combines the total production rate with the physical upper limit of unity for the $R$-hadron signal efficiency.

For BP-S1, the mass gaps $m_{\chi_1^\pm}-m_\chi=4.14$ GeV and $m_{\chi_2^0}-m_\chi=5.20$ GeV give soft visible products. The proper decay lengths $c\tau_{\chi_1^\pm}=0.179$ mm and $c\tau_{\chi_2^0}=0.0301$ mm affect the prompt-search efficiencies. For the specified decays and lifetimes, conservative signal-yield bounds lie below the limits of the searches considered here. This comparison uses leading-order production cross sections, the narrow-width approximation and the adopted approximation for the detector response.

\FloatBarrier
\newlength{\BMoriginalhoffset}\setlength{\BMoriginalhoffset}{\hoffset}
\newlength{\BMoriginalvoffset}\setlength{\BMoriginalvoffset}{\voffset}
\newgeometry{left=16mm,right=16mm,top=16mm,bottom=20mm,footskip=10mm}
\setlength{\hoffset}{0pt}\setlength{\voffset}{0pt}
\clearpage
\BMtable{B0}{B1}{Bino-like benchmark spectra illustrating Higgs-funnel annihilation (BP-B0) and gluino-partner annihilation (BP-B1). Pole masses, $Q_{\rm SUSY}$ and $\mu_{\rm eff}$ are quoted to $0.01$ GeV. LSP fractions have eight significant figures; $\Omega_\chi h^2$ and reaction weights have six decimal places; the remaining numerical entries have three significant figures. The $\mu\to e\gamma$ entries are upper bounds in the stated leading-order flavour treatment.}{tab:benchmark-bino}
\clearpage
\BMtable{S0}{S1}{Singlino-like benchmark spectra illustrating Higgs-funnel annihilation (BP-S0) and electroweakino coannihilation (BP-S1). Numerical precision follows Table~\ref{tab:benchmark-bino}. The $\mu\to e\gamma$ entries are upper bounds in the stated leading-order flavour treatment.}{tab:benchmark-singlino}

\restoregeometry
\setlength{\hoffset}{\BMoriginalhoffset}\setlength{\voffset}{\BMoriginalvoffset}

\FloatBarrier

\section{Conclusions}
\label{sec:conclusions}
We have studied a generalized NMSSM linking the effective Higgs bilinear and the type-II seesaw triplet threshold through a singlet  VEV.   Keeping $\mu_{\rm eff}$ near the soft scale impose correlation between the explicit $\mu$ term and the singlet contribution. The non-renormalization theorem guarantee that large cancellation between the bare $\mu$ term and the term generated from singlet VEV is natural in the supersymmetry limit, although the concrete value of $\mu_{\rm eff}$ also depends slightly on the soft SUSY breaking parameters, as well as finite threshold matching that relate the running parameters to the low-energy spectrum. Integrating out the heavy triplets{, whose masses are determined by the singlet VEV,} yields the Weinberg operator.  Such a scenario can also be UV embedded into a gauge mediation SUSY breaking model, where the triplet that responsible for neutrino mass generation can act as (part of) the messenger that mediate the SUSY breaking effect from the hidden sector to the visible sector.

Under universal inputs for the sfermion masses and trilinear couplings but independent soft inputs for gauginos, the studied spectra separate the high triplet input scale from the neutralino mass. In the  reference relic density range, the bino-like and singlino-like LSP scenarios can constrain the triplet threshold to reach $M_T=4.39\times10^{14}$ and $9.26\times10^{14}$ GeV, respectively. Direct-detection rates reflect the attribute of LSP neutralino and its relic abundance, while nearby superpartners provide additional thermal reactions. BP-B0 and BP-S0 are consistent with heavy-Higgs funnel depletion for bino- and singlino-like neutralinos, respectively. BP-B1 instead illustrates gluino-partner depletion even with a large electroweakino gap, while BP-S1 shows wino coannihilation with a compressed electroweakino spectrum.

Such an extended singlet-triplet framework that connects the neutrino mass generation and $\mu$-term generation has rich collider phenomenology.  With leading-order production cross sections, the conservative bounds on the signal yields of BP-B0 and BP-S0 are below the upper limits from the LHC searches considered here. The bound obtained from the inclusive production rate of BP-B1 is also below these limits.  For the compressed spectrum of BP-S1, the conservative signal-yield bounds are below the limits of the searches considered here for the specified decay modes and lifetimes. This result uses leading-order production cross sections and the narrow-width approximation, together with the adopted approximation for the detector response.  The studied spectra illustrate the coexistence of a high triplet input scale with electroweak-to-TeV neutralino dark matter.

\FloatBarrier
\acknowledgments

This work was supported by the National Natural Science Foundation of China (Grant Nos. 12447167, 12335005, 12275067, 12075213), by Natural Science Foundation of Henan Province (Grant Nos. 242300420656). This work was also supported by the Natural Science Foundation for Distinguished Young Scholars of Henan Province (Grant No. 242300421046).

\end{document}